# Regioselective On-Surface Synthesis of [3]Triangulene Graphene Nanoribbons

Michael C. Daugherty,[†,‡] Peter H. Jacobse,[‡,‡] Jingwei Jiang,[‡,§,‡] Joaquim Jornet-Somoza,[¤,|] Reis Dorit,[†] Ziyi Wang,[‡,§,¶] Jiaming Lu,[‡] Ryan McCurdy,[†] Angel Rubio,[¤,|,¦] Steven G. Louie,[‡,§,*] Michael F. Crommie,[‡,§,¶,*] Felix R. Fischer[†,§,¶,◊,*]

[†]Department of Chemistry, University of California, Berkeley, CA 94720, U.S.A. [‡]Department of Physics, University of California, Berkeley, CA 94720, U.S.A. [§]Materials Sciences Division, Lawrence Berkeley National Laboratory, Berkeley, CA 94720, U.S.A. [¤]Nano-Bio Spectroscopy Group and ETSF, Universidad del País Vasco UPV/EHU, E20018 Donostia, Spain. [|]Max Planck Institute for the Structure and Dynamics of Matter, 22761 Hamburg, Germany. [¦]Center for Computational Quantum Physics (CCQ), The Flatiron Institute, New York, NY 10010, U.S.A. [¶]Kavli Energy NanoSciences Institute at the University of California Berkeley and the Lawrence Berkeley National Laboratory, Berkeley, California 94720, U.S.A. [◊]Bakar Institute of Digital Materials for the Planet, Division of Computing, Data Science, and Society, University of California, Berkeley, CA 94720, U.S.A.

[‡] These authors contributed equally to this work.
* Corresponding Authors

The integration of low-energy states into bottom-up engineered graphene nanoribbons (GNRs) is a robust strategy for realizing materials with tailored electronic band structure for nanoelectronics. Low-energy zero-modes (ZMs) can be introduced into nanographenes (NGs) by creating an imbalance between the two sublattices of graphene. This phenomenon is exemplified by the family of [$n$]triangulenes ($n \in \mathbb{N}$). Here, we demonstrate the synthesis of [3]triangulene-GNRs, a regioregular one-dimensional (1D) chain of [3]triangulenes linked by five-membered rings. Hybridization between ZMs on adjacent [3]triangulenes leads to the emergence of a narrow band gap, $E_{g,exp} \sim 0.7$ eV, and topological end states that are experimentally verified using scanning tunneling spectroscopy (STS). Tight-binding and first-principles density functional theory (DFT) calculations within the local spin density approximation (LSDA) corroborate our experimental observations. Our synthetic design takes advantage of a selective on-surface head-to-tail coupling of monomer building blocks enabling the regioselective synthesis of [3]triangulene-GNRs. Detailed *ab initio* theory provides insight into the mechanism of on-surface radical polymerization, revealing the pivotal role of Au–C bond formation/breakage in driving selectivity.

**Introduction**

Graphene nanoribbons (GNRs) are an emerging class of bottom-up synthesized carbon nanomaterials whose electronic structure can be tailored by the deterministic design of molecular precursors. Laterally confining graphene to a nanoribbon (width < 2 nm) opens a highly tunable band gap that renders these materials attractive candidates for logic devices at the molecular scale.[1-2] More recently, the engineering of low energy states in GNRs has emerged as a robust strategy to induce magnetic ordering in low dimensional phases, superlattices of topologically protected junction states, and even intrinsically metallic band structures in bottom-up synthesized GNRs.[3-7] These advances have been realized by designing structures that imbue nanographenes (NGs) with a sublattice imbalance that gives rise to low-energy states.[8] A sublattice imbalance $\Delta N = N_A - N_B$, where $N_A$ and $N_B$ are the number of carbon atoms on the A and B sublattices, respectively, leads to $\Delta N$ eigenstates at $E = 0$ eV, or ZMs, that are polarized to the majority sublattice.[9] In a chemical picture, these ZMs can be described as the π-radicals associated with open-shell non-Kekule'an structures.[10] The interaction between adjacent ZMs can lead to hybridization and spin correlation effects shaping electronic structure and leading to the emergence of magnetism.[8] Triangular-shaped [$n$]triangulenes are the archetype of sublattice imbalance in NGs and feature a ground-state spin that scales linearly with their size.[11-15] [3]triangulene features an $S = 1$ ground state and interactions between proximal ZMs on neighboring [3]triangulenes linked at their vertices have been studied in dimers,[16] trimers,[17] and 1D spin chains.[18] Interactions between ZMs of triangulenes joined vertex-to-edge (thus connecting the majority and minority sublattices) remain unexplored.

Surface-catalyzed Ullmann-type coupling is a powerful technique for the bottom-up synthesis of [3]triangulene chains and GNRs.[19] The metal surface catalyzes the homolytic cleavage of weak carbon-halogen bonds, facilitating radical step-growth polymerization and subsequent cyclodehydrogenation to form a fully fused aromatic structure. Means of achieving chemoselectivity, however, are limited by the available on-surface polymerization toolkit.[19] Current strategies include exploiting templating effects,[20-23] modulating the composition and structure of the metal surface,[24-26] and leveraging steric hindrance in conjunction with dominant molecular absorption geometries.[3,27] While existing approaches place various constraints on precursor design, perhaps the most restrictive is that to form a regioregular structure, a molecular precursor must be symmetric with respect to a mirror plane perpendicular to the polymerization axis (the *x*-axis in Figure 1A). A synthetic tool for overcoming this requirement in forming regioregular NGs could spur the realization of designer quantum materials exhibiting new magnetic properties and topological phases of matter.[19,28]

Here we report the design and on-surface synthesis of [3]triangulene-GNRs — regioregular [3]triangulene chains featuring fused cyclopentadiene rings. We describe a regioselective surface-catalyzed radical step-growth polymerization between phenyl- and anthracenyl-centered radicals and demonstrate that a regioregular GNR can be formed even though the molecular building block lacks a mirror plane perpendicular to the polymerization axis. The origin of this unusual selectivity is supported by *ab initio* calculations of the on-surface coupling mechanism. STS reveals that arranging [3]triangulene ZMs in a 1D superlattice gives rise to GNRs with a narrow band gap ($E_g$ = 730 meV) and ZM end states. We construct an effective tight-binding (TB) model to describe the GNR electronic structure on the basis of quantum mechanical hopping of electrons between ZMs on adjacent [3]triangulene units and the topological origin of the ZM end states.

**Results and Discussion**

**Synthesis of Molecular Precursors and Surface-Assisted Growth of [3]triangulene-GNRs.** The synthesis of molecular precursor **1** for [3]triangulene GNRs is depicted in Figure 1A. Chemoselective lithiation of 5-bromo-2-iodo-1,3-dimethlylbenzene followed by nucleophilic addition to anthrone gave an intermediate tertiary alcohol.[9] Acid-catalyzed dehydration and rearomatization yielded **2**. Bromination of the 10-anthracenyl position in **2** using *N*-bromosuccinimide (NBS) gave the molecular precursor for [3]triangulene-GNRs, 9-bromo-10-(4-bromo-2,6-dimethylphenyl)anthracene (**1**). Samples of [3]triangulene-GNRs were prepared following established surface-assisted bottom-up GNR growth protocols. Molecular precursor **1** was sublimed in ultra-high vacuum (UHV) from a Knudsen cell evaporator onto a Au(111) surface held at 25 °C. Figure 1B shows a representative topographic STM image of the self-assembled molecular arrangement of precursor **1** into linear structures (Figure S1A). Step-growth polymerization of **1** was induced by annealing the molecule-decorated surface to $T = 200$ °C. Topographic STM images show linear chains of *poly*-**1** localized exclusively along the Au(111) step edges (Figures 1C and S1B,C). A second annealing step at $T = 250$ °C induced both thermal cyclodehydrogenation and radical recombination giving rise to fully fused [3]triangulene-GNRs. Topographic STM images of annealed GNR samples show ribbons with varying degrees of curvature ranging in length from 5 to 20 nm (Figures 1D and S1D–F). Bond-resolved STM (BRSTM) imaging with CO-functionalized tips reveals a structure of [3]triangulene units fused via five-membered rings along the backbone of the GNR (Figure 1E). The relative orientation of each [3]triangulene building block in the GNR backbone suggests that the on-surface polymerization giving rise to *poly*-**1** strongly favors C–C bond formation from a head-to-tail configuration. We herein refer to the *m*-xylyl ring in **1** as the head, and the anthracenyl group as the tail end of the molecule (Figure 1A). The observed head-to-tail (HT) selectivity is remarkable as coupling of the radical intermediate formed from precursor **1** should in principle lead to three discrete geometries. In addition to the observed HT coupling, we would also expect the sterically less encumbered head-to-head (HH) coupling, forming a biphenyl linkage, and tail-to-tail (TT) coupling, giving rise to a bisanthene core, to occur. Despite these alternative possible reaction pathways, the backbone of [3]triangulene-GNRs exclusively features the HT geometry. The only evidence for HH coupling is the observation of dimers localized at the elbow sites of the Au(111) herringbone reconstruction (Figure 1D). The HH reaction intermediate appears to be trapped at the dimer stage and does not react further to form extended oligomers or polymers.

BRSTM images further reveal that the five-membered ring formation between adjacent [3]triangulene units gives rise to patterns of *cis*- and *trans*-linkages that induce local curvature of the ribbon. An apparent preference for the *cis*-conformation, comprising greater than 90% of the linkages, leads to the observed spiral topology of [3]triangulene-GNRs. The proposed *poly*-[3]triangulene intermediate (Figure 1A) could not be trapped on the surface, suggesting that the activation barrier for five-membered ring formation is small. This is further supported by STM images of partially cyclodehydrogenated *poly*-**1** that show fused sections featuring the characteristic curvature induced by the *poly*-[3]triangulene backbone (Figures 1A and S1B,C). Concurrent with the initial cyclodehydrogenation of *poly*-**1**, the emergent triplet π-radical character localized on the C-atoms of the majority sublattice facilitates radical recombination to form additional C–C bonds. Surface-catalyzed dehydrogenation forms the π-conjugated five-membered rings that define the [3]triangulene-GNR backbone. Despite the efficient hybridization of π-radical states of adjacent [3]triangulene units, a finite GNR retains two unpaired electrons (one at either end of the ribbon) that are expected to give rise to characteristic ZMs or localized end states.

**Electronic Structure Characterization of [3]triangulene-GNR.** After elucidating the chemical structure of [3]triangulene-GNRs, we set out to explore their local electronic structure using tunneling spectroscopy (Figures 2 and S2,3). A representative d$I$/d$V$ point spectrum recorded at the position highlighted by the red cross in the topographic STM image (Figure 2A inset) show two prominent features: a broad shoulder at $V_s$ = 400 mV (*State* 1) and a peak at $V_s$ = –330 mV (*State* 2). Constant height d$I$/d$V$ maps of *State* 1 (Figure 2B) show a diffuse striated pattern that closely matches the DFT simulated LDOS map of the LUMO for a cyclic tetradecamer, a proxy for the conduction band (CB) (Figure 2D). Similarly, d$I$/d$V$ maps recorded at $V_s$ = –330 mV (Figure 2C) show a distinctive nodal pattern that closely matches the DFT simulated LDOS map of the corresponding HOMO, a proxy for the valence band (VB) (Figure 2E). The resulting experimental STS band gap of [3]triangulene-GNRs on Au(111) is then $E_{g,exp}$ ~ 0.7 eV.

Differential conductance maps recorded along the center of [3]triangulene-GNRs show the typical signature of a bulk semiconductor — a vanishing DOS at $V_s$ = 0.0 mV (Figure 2G). Both ends of the ribbon exhibit bright nodal features generally associated with low-bias end states. d$I$/d$V$ point spectra recorded over a narrow bias window (–50 mV ≤ $V_s$ ≤ +50 mV) at the positions highlighted by crosses in Figure 2G show sharp peaks centered at $V_s$ = 0.0 mV. These prominent features are characteristic of Kondo resonances arising from the scattering of conduction electrons in the Au substrate by the magnetic moment of localized unpaired electron spins.[29-32] The origin of these ZMs can be traced back to five-membered ring formation along the backbone of [3]triangulene-GNRs. Pairwise π-radical recombination leaves behind a single unpaired electron at either end of the ribbon that manifests as a $S$ = ½ ZM end state localized on the terminating [3]triangulene unit (Figure 1A). This interpretation is further supported by DFT simulated LDOS maps of a finite [3]triangulene-GNR sampled near the Fermi level ($E_F$). The characteristic nodal structure of the ZM end states in d$I$/d$V$ maps is faithfully reproduced by theory at $E = 0.0$ eV (Figure 2H,I).

**First-Principles Electronic Structure Calculation.** We further explored the electronic structure of [3]triangulene-GNRs using *ab initio* density functional theory (DFT). A model of the all-*trans* [3]triangulene-GNR was used in calculations to ensure a periodic unit cell. Figure 3A,B shows the theoretical density of states (DOS) and band structure of [3]triangulene-GNRs calculated using the local spin density approximation (LSDA) for the exchange-correlation potential. The VB and CB are separated by a semiconducting energy gap of $E_{g,\text{DFT}} \sim 0.36$ V. DFT-LSDA typically underestimates band gaps relative to experimental values since it does not account properly for self-energy effects.[33-35] The CB and VB are flanked by sizeable energy gaps that isolate them from the CB+1 and VB–1 (Figure S4).

The low-energy electronic structure of [3]triangulene-GNRs that emerges from *ab initio* DFT can be captured in an effective TB model (Figure 3C). Each triangulene unit bears two ZMs (with on-site energy $\varepsilon_0$) that form the basis states of the TB model. The recombination of two unpaired electrons to form a π-bond leads to the effective hybridization of ZMs on neighboring [3]triangulene units that is described by the hopping term $t_1$. This interaction leads to sublattice mixing which causes the ZMs within a [3]triangulene unit to no longer be orthogonal. The resulting intra-unit interaction is denoted by the hopping term $t_2$. Finally, we consider next-nearest neighbor interactions summarized in the term $t_{\text{NN}}$ which is the arithmetic mean ($t_{\text{NN}} = \frac{1}{2}(t_3 + t_4)$) of the two possible next-nearest neighbor hoppings $t_3$ and $t_4$. The Hamiltonian matrix of this TB model can be expressed as

$$\mathbf{H} = \begin{pmatrix} \varepsilon_0 + 2t_{\text{NN}} \cos ka & t_2 + t_1 e^{ika} \\ t_2 + t_1 e^{-ika} & \varepsilon_0 + 2t_{\text{NN}} \cos ka \end{pmatrix}$$

$$= (\varepsilon_0 + 2t_{NN} \cos ka)\mathbf{I} + \mathbf{H}_{SSH}$$

where $k$ is the electron momentum, $a$ is the lattice constant, and $\mathbf{H}_{SSH}$ the standard Hamiltonian of the form of the Su-Schrieffer-Heeger (SSH) model.[3-4,36] The eigenvalues of this Hamiltonian are

$$E = \varepsilon_0 + 2t_{NN} \cos ka \pm \sqrt{t_1^2 + t_2^2 + 2t_1 t_2 \cos ka}$$

$$= \varepsilon_0 + 2t_{NN} \cos ka \pm E(k)_{SSH}$$

where the positive and negative solutions describe the CB and VB, respectively. This result is identical to the SSH model except for the next-nearest neighbor interaction term, which adds a cosinusoidal modulation to the VB and CB that breaks electron-hole symmetry and accounts for the observed asymmetry between the VB and CB in the band structure (Figure 3A,B). Optimization of the parameters ($\varepsilon_0 = -3.77$ eV, $t_1 = -428$ meV, $t_2 = 106$ meV, $t_{NN} = -145$ meV) provides a good match with the DFT band structure (see Figure S5 for the unfolded band structure).

Provided that $|t_{NN}| < ||t_1| - |t_2||$, the cosine term does not induce a crossing between the VB and CB, and the [3]triangulene-GNR features the same topological properties as the SSH model. When the unit cell is centered on a π-bond between [3]triangulene units, the intra-cell coupling is defined by $t_1$ and the inter-cell coupling by $t_2$. Thus, the intra-cell coupling is larger than the inter-cell coupling, and the [3]triangulene-GNR is topologically trivial. However, when the unit cell is centered on a [3]triangulene, then the intra-cell hopping is defined by $t_2$ and the inter-cell by $t_1$. Thus, inter-cell coupling dominates, and the [3]triangulene-GNR is topologically non-trivial. The GNR explored in Figure 2F,G is terminated by [3]triangulenes units, and the ZMs observed at either end of the ribbon correspond to the topological boundary states predicted by theory.

**Computational Model for the Head-to-Tail Selective On-Surface Polymerization.**

Having resolved the electronic structure of [3]triangulene-GNRs, we return our attention to the unusual head-to-tail regioselectivity observed in the surface assisted radical step-growth polymerization of **1**. In an effort to gain insight into the underlying mechanism that gives rise to this curious selectivity, we performed *ab initio* modeling using the all-electron FHI-aims code.[37] We calculated the activation barriers for the three possible C–C bond formation geometries on a Au(111) surface using DFT at the PBE + vdW + ZORA level (Figure 4A).[37-39] All three modeled reactions are exothermic ranging between $\Delta E = -3.4$ to $-5.8$ eV. The highest transition state energy is predicted for the sterically challenging TT coupling ($E_a = 3.2$ eV), followed by HT ($E_a = 2.0$ eV) and HH ($E_a = 0.6$ eV) coupling. Activation barriers were determined by calculating the energies of relaxed molecular structures along the reaction sequence. The optimized adsorption geometries for the HT coupling pathway are depicted in Figure 4B–I (see Figures S6,7 for the TT and HH coupling pathways). Notably, anthracenyl radical intermediates formed in the thermally induced homolytic cleavage of the C–Br bonds in molecular precursor **1** are stabilized by a covalent interaction with a single Au atom protruding ~0.8–1.0 Å from the plane of the Au(111) surface. The gradual build-up of strain in the coordination of the Au-atom, culminating in Au–C bond breakage in the transition state (TS), provides a major contribution to the activation barrier of the HT and TT reaction profile. While the former mechanism only involves the dissociation of a single Au–C bond in the TS, the latter requires the dissociation of two Au–C bonds to form the TT dimer and accordingly its activation energy is ~1.2 eV higher (Figure S6). Curiously, the calculated activation barrier for the formation of the HH dimer from a pair of phenyl radicals derived from **1** is small and should in principle compete with the HT polymerization.

While the calculated activation barriers qualitatively reproduce the experimental selectivity for the HT over the TT coupling, the absence of HH coupling along the backbone or the ends of [3]triangulene-GNRs requires some additional discussion. A hint can be found in the presence of dimers featuring the HH bonding geometry that adsorb preferentially at the elbow sites of the Au(111) herringbone reconstruction (Figure 1D). Figure S8 shows the relaxed adsorption geometry of the corresponding intermediate at the gold surface. Anthracenyl radicals at either end of the molecule are coordinated to Au atoms of the Au(111) surface. This divalent coordination may represent a kinetic trap that precludes the participation of the HH dimer intermediate in further chain growth before the temperature reaches the threshold of desorption. The annealing of molecule-decorated surfaces to $T = 250$ °C induces a significant decrease in coverage suggesting that small-molecule HH dimers may desorb from the surface before cyclodehydrogenation of [3]triangulene-GNRs is complete. In addition to our hypotheses informed from our mechanistic calculations, a variety of other factors that may influence the observed reaction selectivity are addressed in Supplementary Discussion 1.

**Conclusion**

We have demonstrated rational band engineering using the ZMs of [3]triangulenes assembled vertex-to-edge in a regioregular superlattice. Five-membered ring formation along the GNR backbone facilitates hybridization between ZMs that gives rise to a narrow band gap ($E_{g,exp}$ ~ 0.7 eV) on a Au surface and topological boundary states akin to those described by the SSH model. The chemical bonding, band structure, and ZM end states of [3]triangulene-GNRs are fully characterized with atomic resolution using STM. Tight-binding and first-principles DFT-LSDA calculations support our experimental observations. Our on-surface synthetic strategy shows that a regioregular GNR can be formed even though the molecular precursor does not feature a mirror plane perpendicular to the polymerization axis. Our results provide a framework for the deterministic design of GNR electronic structure using ZMs and offer a new strategy for selective on-surface polymerization.

**ACKNOWLEDGMENT**

This work was primarily funded by the US Department of Energy (DOE), Office of Science, Basic Energy Sciences (BES), Materials Sciences and Engineering Division under contract DE-SC0023105 (molecular synthesis) and contract DE-AC02-05-CH11231 (Nanomachine program KC1203) (STS analysis and DFT calculations). Research was also supported by the National Science Foundation under award CHE-2203911 (molecular design and on-surface growth) and DMR-2325410 (the tight-binding and topological analyses). Part of this research program was generously supported by the Heising-Simons Faculty Fellows Program at UC Berkeley. This research used resources of the National Energy Research Scientific Computing Center (NERSC), a U.S. Department of Energy Office of Science User Facility operated under Contract No. DE-AC02-05CH11231. Computational resources were also provided by the NSF TACC Frontera and NSF through ACCESS resources at the NICS (stampede2). J.J.S. and A.R. acknowledge support from the Cluster of Excellence 'CUI: Advanced Imaging of Matter'- EXC 2056 - project ID 390715994 of the Deutsche Forschungsgemeinschaft (DFG) and UPV/EHU Grupos Consolidados (IT1453-22). We acknowledge support from the Max Planck-New York City Center for Non-Equilibrium Quantum Phenomena. The Flatiron Institute is a division of the Simons Foundation. M.C.D. acknowledges a National Defense Science and Engineering Graduate Fellowship. We thank Dr. Hasan Çelik and the UC Berkeley NMR facility in the College of Chemistry (CoC-NMR) for assistance with spectroscopic characterization. Instruments in the CoC-NMR are supported in part by National Institutes of Health (NIH) award no. S10-OD024998.

**Author Contributions**

All authors have given approval to the final version of the manuscript. M.C.D., P.H.J., and J.J. contributed equally.

**AUTHOR INFORMATION**

**Corresponding Author**

* Felix R. Fischer; Email: ffischer@berkeley.edu

* Steven G. Louie; Email: sglouie@berkeley.edu

* Michael F. Crommie; Email: crommie@berkeley.edu

* Angel Rubio; Email: angel.rubio@mpsd.mpg.de

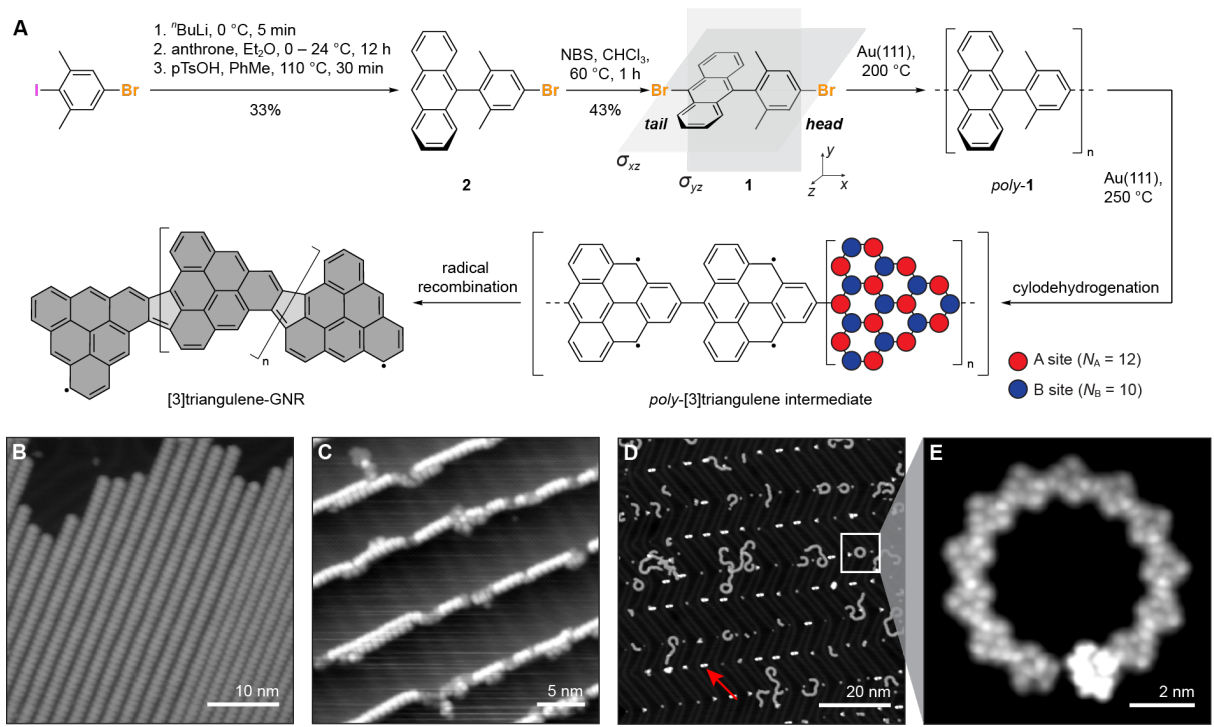

**Figure 1.** Synthesis and structural characterization of [3]triangulene-GNRs. (A) Schematic representation of the bottom-up synthesis of [3]triangulene-GNRs from molecular precursor 1. (B) STM topographic image of a self-assembled island of molecular precursor 1 on Au(111) ($V_s = -2.0$ V, $I_t = 20$ pA). (C) STM topographic image of intermediate linear chains of *poly*-1 following thermal annealing to 200 °C ($V_s = -2.0$ V, $I_t = 30$ pA). (D) STM topographic image of [3]triangulene-GNRs following annealing to 250 °C ($V_s = 1.8$ V, $I_t = 20$ pA). Red arrow: HH dimers at the elbow sites of the Au(111) herringbone reconstruction. (E) Bond-resolved STM (BRSTM) image of the region highlighted by a square in (D) showing the [3]triangulene-GNR structure of regioregular [3]triangulene units fused by five-membered rings ($V_s = 0.0$ V, $V_{ac} = 20$ mV, $f = 533$ Hz).

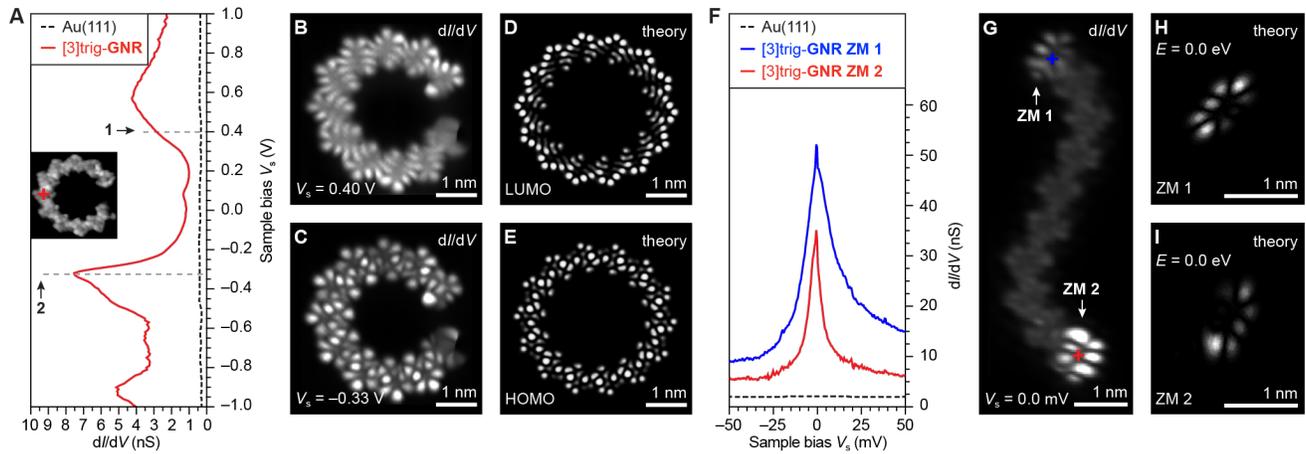

**Figure 2.** Electronic structure of [3]triangulene-GNRs. (A) STS d$I$/d$V$ spectrum recorded on a [3]triangulene-GNR at the position marked by a red cross in the inset BRSTM image (spectroscopy: $V_{ac}$ = 1 mV, $f$ = 533 Hz; imaging: $V_s$ = 0.0 V, $V_{ac}$ = 20 mV, $f$ = 533 Hz). (B,C) Constant height d$I$/d$V$ maps recorded at the indicated biases ($V_{ac}$ = 20 mV, $f$ = 533 Hz). (D,E) DFT simulated LDOS maps of the CB and VB of [3]triangulene-GNRs. A cyclic [3]triangulene-GNR 14-mer was used to reproduce the *cis* geometry of the [3]triangulene-GNR. (F) Low bias d$I$/d$V$ spectra of the two ZMs associated with the [3]triangulene-GNR end states recorded at the positions marked by crosses in (G) (spectroscopy: $V_{ac}$ = 0.5 mV, $f$ = 533 Hz; imaging: $V_s$ = 60 mV, $V_{ac}$ = 20 mV, $f$ = 533 Hz). (G) Constant height d$I$/d$V$ map recorded at $V_s$ = 0.0 mV ($V_{ac}$ = 20 mV, $f$ = 533 Hz). (H,I) DFT simulated LDOS maps ($E$ = 0.0 eV) of the ZMs for the same [3]triangulene-GNR structure shown in (G).

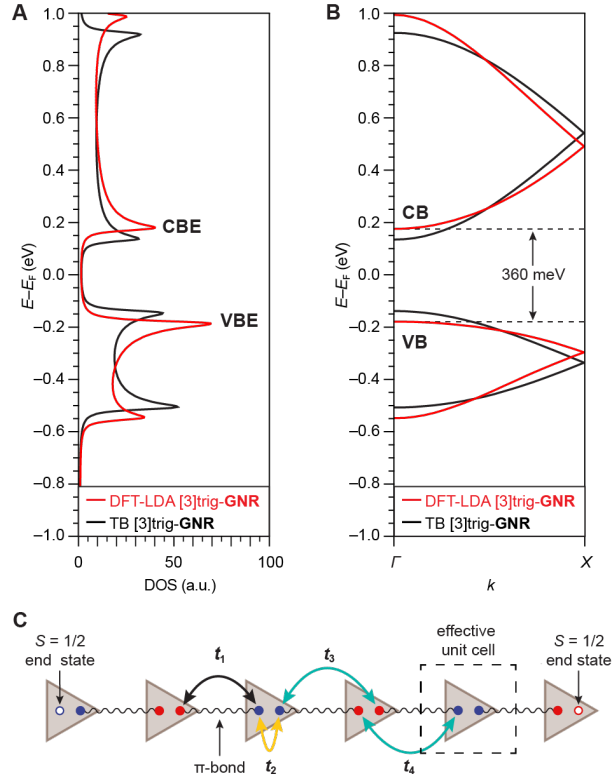

**Figure 3.** First-principle calculations and effective TB model of [3]triangulene-GNRs. (A) DFT-LSDA calculated DOS of [3]triangulene-GNRs showing density increase at band edges characteristic of one-dimensional bands. (B) Band structure of [3]triangulene-GNRs calculated from DFT-LSDA (red) and an effective TB model (black). (C) Effective TB model using basis states that represent the isolated [3]triangulene ZMs coupled via nearest-neighbor electronic hopping parameters $t_1$ (black) and $t_2$ (yellow) and next-nearest-neighbor hopping parameters $t_3$ and $t_4$ (cyan). A ZM at each ribbon end that is weakly-coupled to the other ZMs forms an $S = ½$ end state. Blue and red filled circles represent ZMs on the A and B sublattice, respectively.

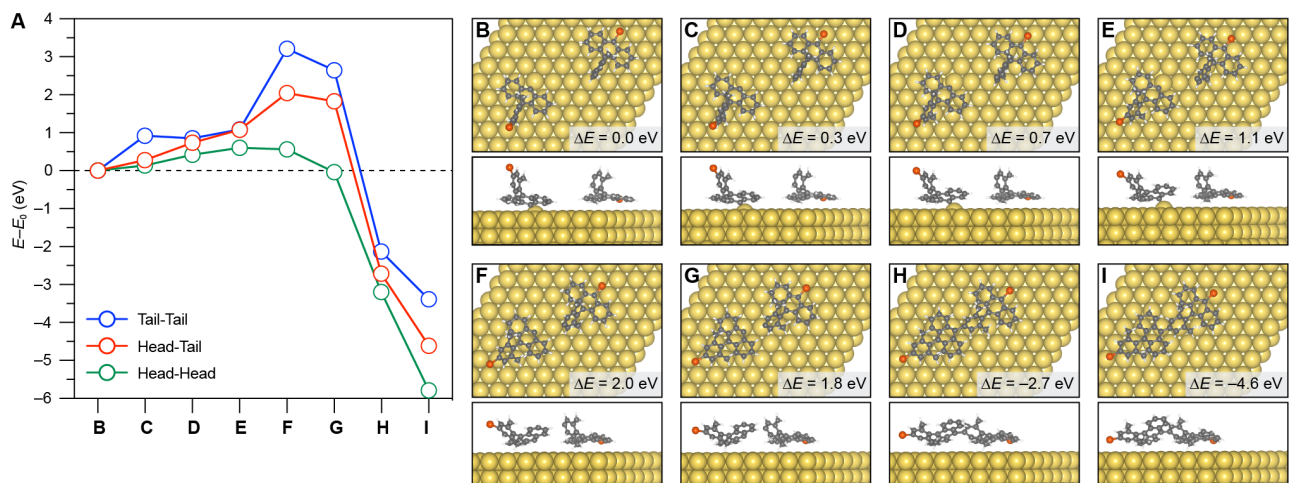

**Figure 4.** Mechanistic calculations of the regioselective coupling of molecular precursor **1** on Au(111). (A) Calculated reaction coordinate diagram showing transition state energies corresponding to the head-to-tail (HT), head-to-head (HH), and tail-to-tail (TT) coupling geometries. (B) Minimized geometries of molecular precursor **1**, (C–H) intermediates, and (I) the product dimer along the head-to-tail coupling pathway.